\documentclass[letterpaper,twocolumn]{jpsj3}
\usepackage{amsfonts}
\usepackage[T1]{fontenc}
\usepackage{txfonts}
\usepackage{amsmath}
\usepackage{amssymb}
\usepackage{graphicx}
\usepackage{color}
\usepackage{tensor}
\usepackage{braket}
\usepackage{bm}
\usepackage{comment}
\usepackage{mathrsfs}
\usepackage{multicol}
\usepackage{cuted}
\usepackage[version=4]{mhchem}
\usepackage[svgnames]{xcolor}
\usepackage{url}
\providecommand{\newblock}{}

\inst{%
$^1$Division of Natural and Environmental Sciences,
  The Open University of Japan, Chiba, 261-8586, Japan \\
$^2$Institute for Molecular Science, Okazaki, Aichi 444-8585, Japan \\
$^3$ Institute of Natural Sciences,
  Ural Federal University, Ekaterinburg 620002, Russia}
\abst{
We present a microscopic theory of chirality-induced orbital selectivity
(CIOS) in helical crystals, in which truly chiral phonons selectively
transfer angular momentum to electronic orbital degrees of freedom.
For a threefold helical crystal with line-group symmetry $L3_1$, we
show that phonon-induced local rotations generate a rotational
electron--phonon interaction proportional to $\hat{L}^{\pm}$, which
drives the orbital transfer $m_{\ell}\to m_{\ell}-m_{s}$ in accordance
with crystal angular momentum (CAM) conservation, where $m_{s}=\pm 1$ denotes the eigenvalue of the phonon rotational mode.
Evaluating $\langle\hat{L}^{z}\rangle$ to leading order in perturbation
theory, we find that the orbital response is suppressed near the $\Gamma$
point and the BZ boundary, and enhanced at intermediate wave
vectors---a feature intimately tied to the degeneracy structure of the
phonon bands.
}

\begin{document}
\title{Microscopic Theory of Chiral-Phonon-Induced Orbital Selectivity
       in Helical Crystals}
\author{Tomomi Tateishi$^{1,2}$,
        Akihito Kato$^{1,2}$,
        Alexander S.\ Ovchinnikov$^{3}$,
        and Jun-ichiro Kishine$^{1,2}$}
\maketitle
Angular momentum transfer between lattice and electronic degrees of freedom 
has recently attracted considerable attention in condensed-matter 
physics.
The idea that phonons carry angular momentum has a long
history~\cite{Vonsovskii1962}, and recent works have renewed interest in
phonon angular momentum and circular or rotational (axial) phonon
modes~\cite{Zhang2014,Zhang2015,Juraschek2025}.
In chiral crystals, however, one must further distinguish such rotational
motion from truly chiral phonons, which require both a rotational
mode of the lattice vibration and a nonzero wave vector with a component
along the rotation axis. In such crystals, the crystal chirality lifts
the degeneracy of phonon dispersions resolved by
CAM~\cite{Hamada-Minamitani-Murakami2018,
Ishito2022,Ishito2023,Ueda2023,oishi2024,Togawa2024}.
This viewpoint has led to extensive studies of phonons with true chirality in the sense of Barron~\cite{Barron1982}, their chirality-resolved dispersions, and related responses in chiral crystals~\cite{Kishine2020,Tsunetsugu2023,Kato2023,Yao2025}.

Helical crystals with monoaxial screw symmetry provide a particularly
suitable platform for this problem, since their elementary excitations
are classified by irreducible representations of the line
group~\cite{Bozovic1978,Bozovic1981,Bozovic1984}.
Within this framework, electrons and phonons share the same screw
symmetry and are both characterized by crystal momentum and
CAM~\cite{Kato2023,TKK2025}---a common symmetry footing that enables
direct CAM exchange between the two subsystems.
In helical crystals, ionic rotational motion, orbital character, and chirality
are thus intimately intertwined.
In this context, several theoretical routes toward this angular momentum
transfer have been proposed~\cite{Funato2024,Yao2025,Yokoyama2025,SatoKato2025,Tsirkin2018,YaoMurakami2022},
and experimental orbital-current accumulation has also been
reported~\cite{Nabei2026}.

The orbital angular momentum (OAM) carried by chiral phonons, once
transferred to the electronic orbital sector, serves as a key
intermediate toward spin polarization via spin--orbit coupling---a
route that may provide a microscopic account of chirality-induced
spin selectivity (CISS), the generation of spin-polarized transport
by structural chirality without magnetic
constituents~\cite{Ray1999,Gohler2011}, whose microscopic mechanism
remains a subject of active debate. While the role of orbital
polarization within the electron subsystem has been
pointed out~\cite{Yang2025}, the electron--phonon coupling channel
by which chiral phonons transfer angular momentum to the electronic
orbital sector remains unexplored.

\begin{figure}[ptb]
\begin{center}
\includegraphics[width=0.9\columnwidth]{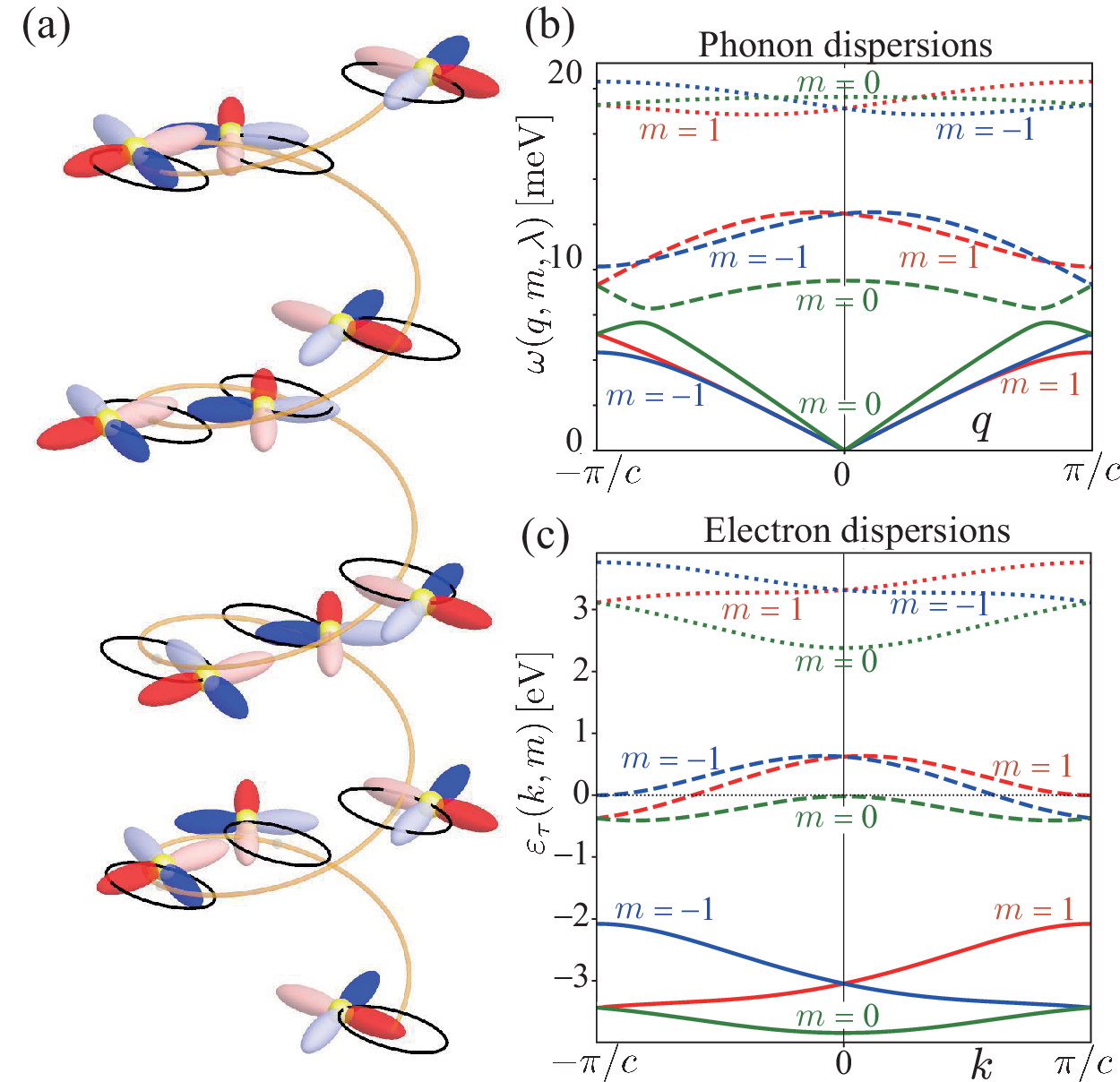}
\end{center}
\caption{%
(a) Schematic illustration of a rotational phonon mode in a helical crystal and the $p$-orbital rotation comoving with the displaced ions. The yellow spheres denote ions undergoing rotational motion, the black ellipses represent their classical orbits, and the red and blue lobes depict the two planar electronic $p$ orbitals following the ionic displacement. 
(b) Phonon and (c) electron dispersions for a chiral crystal belonging to  helical line-group symmetry $L3_{1}$, resolved by CAM $m=\pm1,0$.
Electron dispersions are obtained using the Slater--Koster parameters:
$t^{0}_{++}=t^{0}_{--}=0.242964$, $t^{0}_{00}=1.034072$,
$t^{0}_{+-}=-0.496482+0.859932i$,
$t^{0}_{+0}=-0.665492-1.152666i$,
$t^{0}_{-0}=(t^{0}_{+0})^{*}$,
and $E_{\pm1}=0.15$, $E_0=-0.20$.
All parameters are in eV and taken from those of tellurium (Te) \cite{harrison1989electronic} for illustrative purposes.
}
\label{Fig1}
\end{figure}

In Ref.~[\citen{TKK2025}], the coupling in a threefold helical crystal was formulated from irreducible representations of the helical line group $L3_1$, but only for $s$-orbital electrons. Here we extend it to $p$-orbital electrons ($m_{\ell}=0,\pm1$)~\cite{Hu2024}, and derive the rotational electron--phonon vertex analytically, thereby establishing a microscopic symmetry-based theory of CIOS. 
We confirm it by evaluating its contribution to $\langle\hat{L}^{z}\rangle$.
As an effective one-dimensional model of a three-dimensional helical crystal, we retain electron and phonon modes along the helical axis and classify them by $L3_{1}$ symmetry. We follow the formulation of Ref.~[\citen{TKK2025}] and focus on the $p$-orbital effects. Figure~\ref{Fig1}(a) schematically shows the rotational phonon mode and co-rotating $p$ orbitals.

The lattice sites along the helix are indexed by $l=1,2,\ldots,3N$, where $N$ is the number of unit cells.
The displaced atomic position is written as $\bm{r}_{l}(t)=\bm{R}_{l}+\bm{u}_{l}(t)$,
where $\bm{R}_{l}$ is the equilibrium position and
$\bm{u}_{l}(t)=\sum_{s=\pm,0} u_{l}^{s}(t)\,\bm{e}^{s}$
denotes the nuclear displacement.
Here, the vectors
$\bm{e}^{\pm}:=(\bm{e}^{1}\mp i\bm{e}^{2})/\sqrt{2}$ and
$\bm{e}^{0}:=\bm{e}^{3}$
form a chiral basis.
The phonon eigenstates are labeled by the crystal momentum
$q_n = 2\pi n/(cN)$ \,$(n=0,1,\ldots,N-1)$, the CAM $m=0,\pm1$, and the branch index $\lambda$, where $c$ is the lattice constant.
The phonon dispersion is shown in Fig.~\ref{Fig1}(b).

The quantized phonon field is then written as~\cite{TKK2025}
\begin{align}
u_{l}^{s}(t)
&=\frac{1}{\sqrt{3N}}
\sum_{n=0}^{N-1}\sum_{m=0,\pm 1}
e^{-i(l-1)\left[q_{n}\frac{c}{3}+\alpha(m-m_{s})\right]}
\hat{w}_{nm}^{s}(t), \label{uls}
\\
\hat{w}_{nm}^{s}(t)
&=\sum_{\lambda=1}^{3}
\left(\frac{\hbar}{2M\omega_{nm}^{\lambda}}\right)^{\!\frac{1}{2}}
\!\left(\hat{a}_{n,m}^{\lambda}(t)+\hat{a}_{-n,-m}^{\lambda\dagger}(t)\right)
v_{nm}^{\lambda(s)},
\label{wnms}
\end{align}
where $\omega_{nm}^{\lambda}$ and $v_{nm}^{\lambda(s)}$ are the eigenvalues and eigenvectors of the phonon dynamical matrix, respectively, and $m_s$ is the eigenvalue of the threefold rotation in the chiral basis. Here, $M$ denotes the ionic mass, and $\hat a_{n,m}^{\lambda}(t)$ is the phonon annihilation operator.

We now turn to the electronic sector.
Let $\phi_{k}$ ($k=x,y,z$) denote the Cartesian $p$ orbitals.
We introduce the angular-momentum eigenstates
$\ket{\phi_{\pm}}=(\ket{\phi_{x}}\pm i\ket{\phi_{y}})/\sqrt{2}$
and
$\ket{\phi_{0}}=\ket{\phi_{z}}$,
satisfying
$\hat{L}^{z}\ket{\phi_{m_{\ell}}}=\hbar m_{\ell}\ket{\phi_{m_{\ell}}}$
with $m_\ell=\pm1,0$, where $m_\ell$ labels the orbital basis indices.
Let $\hat{c}_{l,m_{\ell}}$ be the annihilation operator of a
$p$-orbital electron.
Following Refs.~[\citen{Bozovic1984,Kato2023,TKK2025}], its Fourier transform is
\begin{align}
\hat{c}_{l,m_{\ell}}
=
\frac{1}{\sqrt{3N}}
\sum_{n,m}
e^{-i(l-1)\left[k_{n}\frac{c}{3}+\alpha(m+m_{\ell})\right]}
\hat{c}_{n,m,m_{\ell}},
\label{canihi}
\end{align}
with $k_{n}=2\pi n/(cN)$.
Using Eq.~(\ref{canihi}), the tight-binding Hamiltonian becomes
\begin{align}
\hat{\mathcal{H}}_{\mathrm{el}}
&=\sum_{n,m}\sum_{m_{\ell},m_{\ell'}}
(\mathbf{B}_{nm})_{m_{\ell}m_{\ell'}}
\hat{c}^{\dagger}_{n,m,m_{\ell}}
\hat{c}_{n,m,m_{\ell'}},
\\
(\mathbf{B}_{nm})_{m_{\ell}m_{\ell'}}
&=
\delta_{m_{\ell}m_{\ell'}}E_{m_{\ell}}
-t^{0}_{m_{\ell}m_{\ell'}}
\!\left(
e^{i\left(k_{n}\frac{c}{3}+\alpha(m+m_{\ell})\right)}
+
e^{-i\left(k_{n}\frac{c}{3}+\alpha(m+m_{\ell'})\right)}
\right),
\end{align}
where $E_{m_{\ell}}$ and $t^{0}_{m_{\ell}m_{\ell'}}$ are the on-site energies and equilibrium Slater--Koster hopping integrals (SKHIs).
Diagonalization gives
$\mathbf{B}_{nm}\ket{\phi_{n,m,\tau}}=\varepsilon_{nm}^{\tau}\ket{\phi_{n,m,\tau}}$, 
where $\tau=1,2,3$ is the band index replacing $m_{\ell}$.
The bands are shown in Fig.~\ref{Fig1}(c).
The two bases are related by
$\ket{\phi_{n,m,\tau}}=\sum_{m_{\ell}}\ket{\phi_{n,m,m_{\ell}}}\braket{\phi_{n,m,m_{\ell}}|\phi_{n,m,\tau}}
$.

From the character of $L3_{1}$,\,
$\chi_{nm}(p)=e^{i(k_nc/3+m\alpha)p}\, (p=0,1,\ldots,3N-1)$\cite{TKK2025},
the condition $\chi_{nm}^{*}(p)=\chi_{nm'}(p)$ ($m\neq m'$) is satisfied only at $\Gamma$ and the BZ boundary ($k=\pm\pi/c$), giving the time-reversal pairs $(+1,-1)$ and $(0,\mp1)$, respectively.
The branch participating in this sticking thus switches between $\Gamma$ and the boundary, while the remaining one stays isolated. This sticking pattern is shared by the phonon and electron bands in Figs.~\ref{Fig1}(b) and (c).

We next examine how the $p$-orbital operators transform under rotation;
the electron spin plays no role here and is suppressed.
The OAM operator at site $l$ is written by
\begin{align}
\hat{\mathcal{L}}^{k}_{l}
=
\sum_{m_{\ell},m_{\ell'}}
\hat{c}^{(\mathrm{lab})\dagger}_{l,m_{\ell}}
(L^{k})_{m_{\ell},m_{\ell'}}
\hat{c}^{(\mathrm{lab})}_{l,m_{\ell'}},
\end{align}
where $\hat{c}^{(\mathrm{lab})}_{l,m_{\ell}}$ and $\ket{\phi_{l,m_{\ell}}}$
denote the annihilation operator and state in the laboratory frame, and
$(L^{k})_{m_{\ell},m_{\ell'}}:=\bra{\phi_{l,m_{\ell}}}\hat{L^{k}}\ket{\phi_{l,m_{\ell'}}}$.
Under the rotation $\hat{U}_{l}=\exp[-\frac{i}{\hbar}\bm{\theta}_{l}
\cdot\hat{\bm{\mathcal{L}}}_{l}]$, the laboratory and local fermionic
operators are related by
\begin{equation}
\hat{c}^{(\mathrm{lab})}_{l,m_{\ell}}
=
\sum_{m_{\ell'}}
\left[\exp\!\left(\tfrac{i}{\hbar}\bm{\theta}_l\cdot\bm{L}\right)
\right]_{m_{\ell}m_{\ell'}}
\hat{c}^{(\mathrm{loc})}_{l,m_{\ell'}}.
\end{equation}
We first take $\bm{\theta}_l=\alpha(l-1)\bm{e}_z$ with $\alpha=2\pi/3$ for
the global helical rotation, then introduce the additional phonon-induced
local rotation $\delta\bm{\theta}_l$.
The helical lattice admits no natural continuum limit, since taking the helix radius to zero collapses the structure and changes its symmetry. We therefore define the local rotation vector:
\begin{equation}
\delta\bm{\theta}_l
\simeq
\frac{3}{2c}\bm{e}^{3}\times(\bm{u}_{l+1}-\bm{u}_{l}),
\label{dtheta}
\end{equation}
directly on the discrete lattice, as the rotation induced by the relative displacement of neighboring sites along the helix:
$\partial^{j}u^{i}_{l}\simeq (u^{i}_{l+1}-u^{i}_{l})/(c/3)\,\delta_{j3}$.
This construction is not a long-wavelength approximation; rather, it incorporates the full finite-lattice structure and remains valid at all wave vectors up to the Brillouin-zone (BZ) boundary. For reference, in continuum elasticity the analogous quantity is $\bm{\delta\theta}=\frac{1}{2}\nabla\times\bm{u}$.
Because $\delta\bm{\theta}_l\perp\bm{e}^3$, the induced rotation
couples exclusively to the in-plane orbital operators $\hat{L}^{(\pm)}$, as
will be seen explicitly below.
Expanding to first order in $\delta\bm{\theta}_l$, we obtain
\begin{align}
\hat{c}^{(\mathrm{lab})}_{l,m_{\ell}}
\simeq
e^{i(l-1)m_{\ell}\alpha}
\sum_{m_{\ell'}}
\!\left[
\delta_{m_{\ell}m_{\ell'}}
+\frac{i}{\hbar}\delta\bm{\theta}_{l}\cdot(\bm{L})_{m_{\ell}m_{\ell'}}
\right]
\hat{c}^{(\mathrm{loc})}_{l,m_{\ell'}}.
\label{labtoloc}
\end{align}

To describe the $p$-electron--phonon interaction, we use the modified
tight-binding approximation~\cite{Friedel1970,Barisic1972I,
TKK2025} in the laboratory frame,
\begin{align}
\hat{\mathcal{H}}
= -\sum_{\langle l,l'\rangle}\sum_{m_{\ell},m_{\ell'}}
\hat{c}^{(\mathrm{lab})\dagger}_{l,m_{\ell}}\,
t_{m_{\ell},m_{\ell'}}(\bm{r}_l-\bm{r}_{l'})\,
\hat{c}^{(\mathrm{lab})}_{l',m_{\ell'}}.
\label{htight}
\end{align}
Expanding the overlap integral about equilibrium positions and substituting
Eq.~(\ref{labtoloc}) into Eq.~(\ref{htight}), retaining terms to first
order in $\bm{u}_l$ (including the rotation $\delta\bm{\theta}_l$), yields
\begin{equation}
\hat{\mathcal{H}}_{\mathrm{el\text{-}ph}}
=\hat{\mathcal{H}}^{0}_{\mathrm{el\text{-}ph}}
+\hat{\mathcal{H}}^{(\mathrm{rot})}_{\mathrm{el\text{-}ph}}.
\end{equation}

Here, $\hat{\mathcal{H}}^{0}_{\mathrm{el\text{-}ph}}$ and
$\hat{\mathcal{H}}^{(\mathrm{rot})}_{\mathrm{el\text{-}ph}}$
originate from the first and second terms of Eq.~(9), respectively.
The explicit form of
$\hat{\mathcal{H}}^{0}_{\mathrm{el\text{-}ph}}$
was given in Ref.~[\citen{TKK2025}].
In contrast,
$\hat{\mathcal{H}}^{(\mathrm{rot})}_{\mathrm{el\text{-}ph}}$
couples directly to the rotational phonon mode and explicitly contains
$\hat{L}$, which reads
\begin{align}
\hat{\mathcal{H}}^{(\mathrm{rot})}_{\mathrm{el\text{-}ph}}
=
&-\frac{i}{\hbar}
\sum_{\langle l,l'\rangle}
\sum_{m_{\ell},m_{\ell'}}
\sum_{\bar{m}_{\ell},\bar{m}_{\ell'}}
e^{-i(l-1)m_{\ell}\alpha}
e^{i(l'-1)m_{\ell'}\alpha}
\notag\\
&\times\left\{
\delta_{m_{\ell'}\bar{m}_{\ell'}}
\hat{c}^{(\mathrm{loc})\dagger}_{l,\bar{m}_{\ell}}
(\delta\bm{\theta}_l\cdot\bm{L})_{\bar{m}_{\ell}m_{\ell}}
t^{0}_{m_{\ell}m_{\ell'}}
\hat{c}^{(\mathrm{loc})}_{l',\bar{m}_{\ell'}}
\right.
\notag\\
&\left.
-\delta_{m_{\ell}\bar{m}_{\ell}}
\hat{c}^{(\mathrm{loc})\dagger}_{l,\bar{m}_{\ell}}
t^{0}_{m_{\ell}m_{\ell'}}
(\delta\bm{\theta}_{l'}\cdot\bm{L})_{m_{\ell'}\bar{m}_{\ell'}}
\hat{c}^{(\mathrm{loc})}_{l',\bar{m}_{\ell'}}
\right\},
\label{helph}
\end{align}
where $t^{0}:=t(\bm{R}_{l}-\bm{R}_{l'})$.
Since the SKHIs, $t_{m_\ell,m_{\ell'}}^{0}$, for helical $p$-orbital electrons already incorporate the helical geometry and are non-diagonal\cite{Slater1954}, both $\hat{\mathcal{H}}^{0}_{\mathrm{el\text{-}ph}}$ and $\hat{\mathcal{H}}^{(\mathrm{rot})}_{\mathrm{el\text{-}ph}}$ mix $m_{\ell}$. However, since the transfer of phonon angular momentum to the electronic OAM is enabled only by
$\hat{\mathcal{H}}^{(\mathrm{rot})}_{\mathrm{el\text{-}ph}}$,
we focus on this term, which is entirely absent for $s$ electrons.
Using Eqs.~(\ref{uls}), (\ref{wnms}), and the relation
$\delta\bm{\theta}_l\cdot\hat{\bm{L}}
=\hat{L}^{(-)}\delta\theta^{(+)}_l+\hat{L}^{(+)}\delta\theta^{(-)}_l$,
together with the orthogonality relation
$\sum_{l=1}^{3N}e^{-i(l-1)\{(q_n-q_{n'})\frac{c}{3}+\alpha(m-m')\}}
=3N\delta_{n,n'}\delta_{m,m'}$
and the ladder identity
$(L^{(\pm)})_{m_{\ell'}m_{\ell}}=\sqrt{2}\hbar\,\delta_{m_{\ell'},m_{\ell}\pm 1}$,
we evaluate Eq.~(\ref{helph}) and obtain
\begin{align}
\hat{\mathcal{H}}^{(\mathrm{rot})}_{\mathrm{el\text{-}ph}}
&=
\sum_{n,m,\lambda}\sum_{n',m'}\sum_{s=\pm}
g^{\lambda s}_{nm}\,
\mathcal{T}^{s}_{n,m;n',m'}
\left(\hat{a}_{n,m}^{\lambda}+\hat{a}^{\lambda\dagger}_{-n,-m}\right),
\label{elphorot}
\end{align}
where
\begin{align}
g^{\lambda s}_{nm}
&:=
-\frac{6i}{c\sqrt{3N}}
\left(\frac{\hbar}{M\omega^{\lambda}_{nm}}\right)^{\!\frac{1}{2}}
v_{nm}^{\lambda(s)}\,
e^{-\frac{i}{2}\left[q_n\frac{c}{3}+\alpha(m-m_s)\right]}
\notag\\
&\quad\times
\sin\!\left[\tfrac{1}{2}\!\left(q_n\tfrac{c}{3}+\alpha(m-m_s)\right)\right],
\label{glambdas}
\\
\mathcal{T}^{s}_{n,m;n',m'}
&:=
\sum_{m_{\ell},m_{\ell'}}
\Biggl\{
\cos\!\left(k_{n'}\tfrac{c}{3}+\alpha m'\right)
\notag\\
&\qquad\times
\hat{c}^{\dagger}_{n+n',m+m',m_{\ell}-m_s}
t^0_{m_{\ell},m_{\ell'}}
\hat{c}_{n',m',m_{\ell'}}
\notag\\
&\quad
-\cos\!\left[(q_n+k_{n'})\tfrac{c}{3}+\alpha(m+m')\right]
\notag\\
&\qquad\times
\hat{c}^{\dagger}_{n+n',m+m',m_{\ell}}
t^0_{m_{\ell},m_{\ell'}-m_s}
\hat{c}_{n',m',m_{\ell'}}
\Biggr\}.
\label{Ts}
\end{align}
Here, $s=\pm$ labels the chiral components, with $m_s\equiv s$
(i.e., $m_{+}=+1$, $m_{-}=-1$).
Equations~(\ref{elphorot})--(\ref{Ts}) constitute the key result of this study.
The action of $\hat{L}^{(\pm)}$ induces an OAM transfer
$m_{\ell}\to m_{\ell}-m_{s}$ in the electronic sector, appearing in both
the creation operator and the hopping matrix elements.
This transfer directly reflects the coupling between the rotational mode specific to chiral phonons and the intrinsic rotational character of electronic $p$-orbitals, mediated by the operator $\hat{U}_{l}$. 
It should be noted that the mechanism requires partially filled $p$-orbitals 
near the Fermi level; a fully occupied shell Pauli-blocks the rotational 
vertex $\hat{\mathcal{H}}^{(\mathrm{rot})}_{\mathrm{el\text{-}ph}}$,
suppressing the $m_\ell \to m_\ell - m_s$ transfer. By extension, partial 
orbital filling near the Fermi level may constitute a general prerequisite 
for efficient angular-momentum transfer across CISS-active materials, 
whether the relevant states are of $p$- or $d$-orbital character.
Furthermore, this interaction reflects the conservation of both the crystal momentum $q_n$ and the CAM $m$, and is invariant under time-reversal symmetry, consistent with Ref.~[\citen{TKK2025}].
In this regard, it is noteworthy that the direct coupling of phonon mechanical angular momentum\cite{Zhang2014, TKK2025} to electronic degrees of freedom was also recently established by two of the present authors~\cite{kato2026Effective}.

\begin{figure}[b]
\begin{center}
\includegraphics[width=6.5cm]{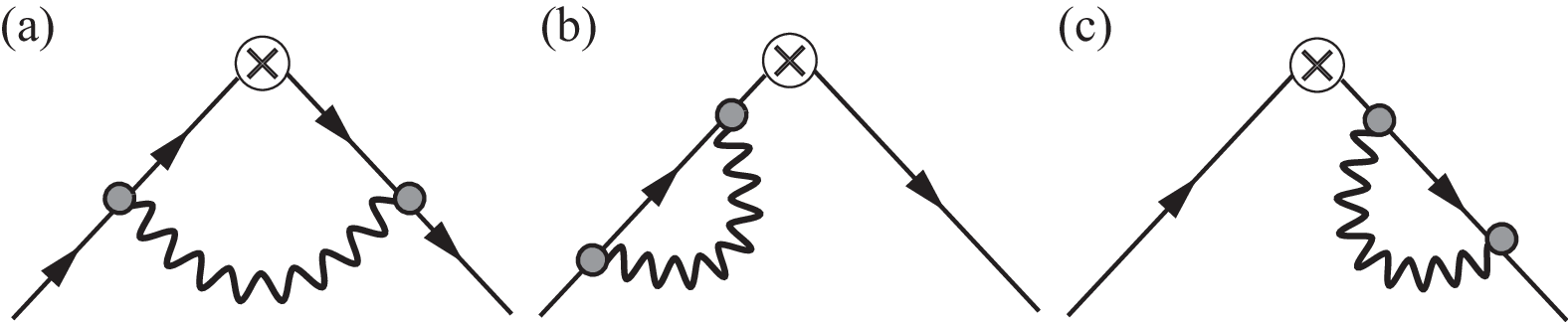}
\end{center}
\caption{
Diagrammatic representation of the leading-order contributions to 
$\langle\hat{L}^{z}\rangle$ from the electron--phonon interaction. Solid 
arrowed lines: electron propagation; wavy lines: chiral-phonon propagation; 
filled circles: rotational electron--phonon vertices at which CAM is exchanged 
between electrons and phonons (the phonon CAM is correlated with its mechanical 
angular momentum\cite{TKK2025}); crossed circles: insertion of $\hat{L}^{z}$. 
Process (a) shows the dominant contribution to $\langle\hat{L}^{z}\rangle$ 
evaluated in this work; processes (b) and (c) show self-energy corrections to the electron propagation.}
\label{fig_Lz-diagram}
\end{figure}

\begin{figure*}[tb]
\begin{center}
\includegraphics[width=0.8\textwidth]{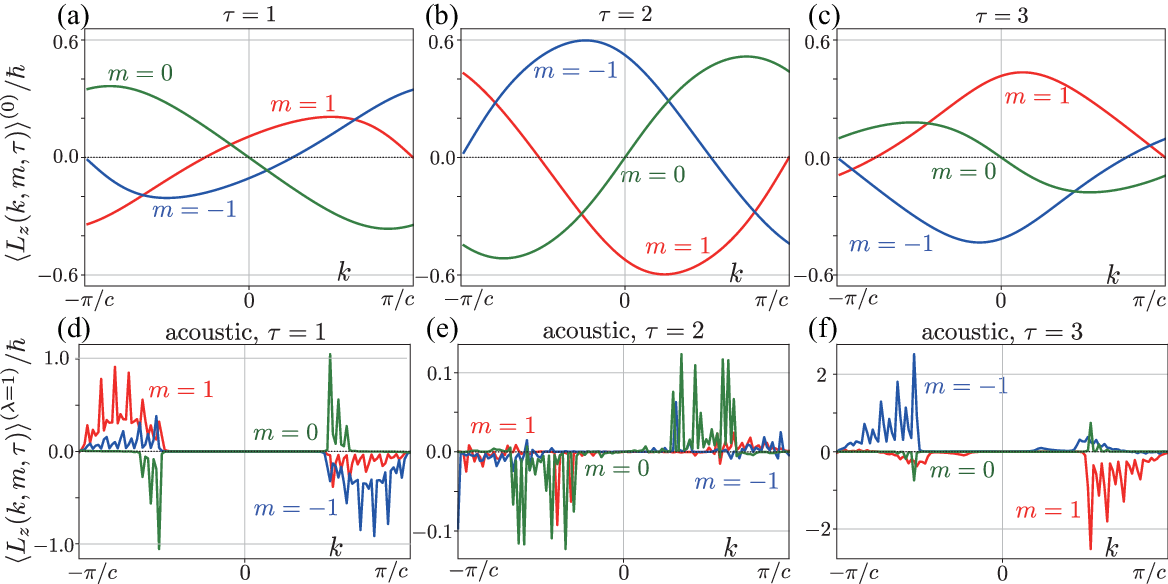}
\end{center}
\caption{
Expectation value of $\langle \hat{L}^{z}\rangle/\hbar$ as a function of $k$.
The same Slater--Koster parameters as in Fig.~\ref{Fig1} (c) are used.
The value $1/(Mc^{2}) = 1.0 \times 10^{42}~\mathrm{kg}^{-1}\mathrm{m}^{-2}$, 
consistent with typical ionic masses and lattice constants 
of helical crystals, is adopted.
The top row (a)--(c) shows the unperturbed contribution for different initial electronic states, $\tau=1,2,3$, while the lower panels (d)--(f) display the perturbative contribution [second term in Eq.~(\ref{Lz1st})] for acoustic phonon branch $\lambda=1$. Within each panel, curves represent the $m=+1$ (red), $m=-1$ (blue), and $m=0$ (green) branches.}
\label{fig_Lz}
\end{figure*}

We have now established the electron-phonon coupling, we compute the expectation value of $L^{z}$ to leading order in perturbation theory.
The validity of the perturbative expansion is guaranteed by
$(\hbar/M\omega_{nm}^{\lambda})^{1/2}/c\sim 10^{-2}\ll 1$,
confirming that the electron--phonon coupling is in the weak-coupling
regime.
Up to second order in the interaction, three processes contribute, as
diagrammatically shown in Fig.~\ref{fig_Lz-diagram}.
Process (a) represents the dominant contribution to $\langle\hat{L}^{z}\rangle$, 
in which the electronic state is dressed by the emission and reabsorption of a 
single virtual chiral phonon via $\hat{\mathcal{H}}^{(\mathrm{rot})}_{\mathrm{el\text{-}ph}}$, 
and the resulting change in OAM is evaluated by the insertion 
of $\hat{L}^{z}$. Processes (b) and (c), by contrast, enter only indirectly 
through self-energy renormalization and are therefore ignored below.
Taking $\hat{\mathcal{H}}^{(\mathrm{rot})}_{\mathrm{el\text{-}ph}}$ as the
perturbation to the unperturbed state $\ket{\phi_{k_{n'},m',\tau'};0}$,
comprising 
an electronic state $|\phi_{k_{n'},m',\tau'}\rangle$ and the phonon vacuum $|n_{q_n,m,\lambda}=0\rangle$,
the expectation value of $\langle \hat{L}^{z}\rangle$ to leading order is
\begin{align}
\langle \hat{L}^{z}\rangle
&=
\bra{\phi_{k_{n'},m',\tau'}}\hat{L}^{z}\ket{\phi_{k_{n'},m',\tau'}}
\notag\\
&+
\sum_{q_n,m,\lambda}\sum_{\tau,\tau''}
\mathcal{L}^{\tau\tau''}_{q_n m}
\frac{
\mathcal{M}^{\tau}_{q_n,m,\lambda}
\left(\mathcal{M}^{\tau''}_{q_n,m,\lambda}\right)^{*}
}{
\!\left(\Delta^{\tau}_{q_n m\lambda}+i\hbar/\tau_{\mathrm{qp}}\right)
\!\left(\Delta^{\tau''}_{q_n m\lambda}-i\hbar/\tau_{\mathrm{qp}}\right)},
\label{Lz1st}
\end{align}
where
$\mathcal{L}^{\tau\tau''}_{q_n m}$
$
:=\bra{\phi_{k_{n'}-q_n,m'-m,\tau}}\hat{L}^{z}
\ket{\phi_{k_{n'}-q_n,m'-m,\tau''}}$,
$\Delta^{\tau}_{q_n m\lambda}$
$:=\varepsilon_{k_{n'},m'}^{\tau'}
-\varepsilon_{k_{n'}-q_n,m'-m}^{\tau}-\omega_{nm}^{\lambda}$,
and
$\mathcal{M}^{\tau}_{q_n,m,\lambda}$
$
:=\bra{\phi_{k_{n'}-q_n,m'-m,\tau};1_{q_n,m,\lambda}}
\hat{\mathcal{H}}^{(\mathrm{rot})}_{\mathrm{el\text{-}ph}}
\ket{\phi_{k_{n'},m',\tau'};0}$.
The broadening $\eta:=\hbar/\tau_{\mathrm{qp}}=1~\mathrm{meV}$
is a phenomenological regularization parameter physically
corresponding to the quasi-particle inverse lifetime due to
electron--phonon scattering; the qualitative features in
Fig.~\ref{fig_Lz} remain unchanged for $\eta$ in the range
$0.1$--$10~\mathrm{meV}$, which covers typical values in
crystalline solids.

The unperturbed expectation value is
$\bra{\phi_{k_{n},m,\tau}}\hat{L}^{z}\ket{\phi_{k_{n},m,\tau}}
=\hbar\sum_{m_{\ell}}|\braket{\phi_{n,m,m_{\ell}}|\phi_{n,m,\tau}}|^{2}m_{\ell}$.
Figs.~\ref{fig_Lz}(a)-(c) show its $k$ dependence for $\tau=1,2,3$.
Since the symmetry-adapted eigenstates are not eigenstates of $\hat{L}^z$, the unperturbed contribution is already nonzero, reflecting the static orbital polarization of the chiral band structure, while the perturbative contribution [Figs.~3(d)--(f)] represents the dynamical, CAM-selective transfer from chiral phonons to electrons.

The $k$ dependence of the perturbative contribution [second term in 
Eq.~(\ref{Lz1st})] is shown in Figs.~\ref{fig_Lz}(d)--(f) for acoustic phonons,
which dominate phonon transport.
The response is suppressed near the $\Gamma$ point and at the BZ boundary 
due to phonon band degeneracies ---the time-reversal sticking 
identified through the character $\chi_{nm}(p)$ pairs branches with 
opposite OAM, whose contributions cancel upon degeneracy--- and is enhanced 
in the intermediate region.
Near the $\Gamma$ point, the phonon CAM $m$ is closely associated with 
longitudinal/transverse modes, while away from $\Gamma$ these modes become mixed.
Since the present electron--phonon interaction originates from the rotational 
(transverse) phonon modes, this mixing leads to a stronger interaction at 
intermediate wave vectors.
We have confirmed that optical modes exhibit qualitatively the same behavior 
and do not present those results here.
Furthermore, Figs.~\ref{fig_Lz}(d)--(f) exhibit numerous peaks, originating 
not only from the crystal momenta $q_n$ and $k_{n'}$ but also from the large 
number ($3^{5}$) of combinations of discrete indices $(m,\lambda,m',\tau,\tau'')$,
so that many resonant peaks appear even for electrons propagating along the helical axis.

We briefly discuss how the results shown in Fig.~\ref{fig_Lz} transform under chirality inversion.
Consider the left-handed state $\ket{\phi_{k_{n},m,\tau}}_{L}$ in the Hamiltonian with $L3_2$ symmetry, and compare it with the corresponding right-handed state $\ket{\phi_{k_{n},m,\tau}}_{R}$ obtained here.
By inserting the inversion operator and its inverse, one finds
${}_{L}\!\bra{\phi_{k_{n},m,\tau}}\hat{L}^{z}\ket{\phi_{k_{n},m,\tau}}_{L}
=
-\,{}_{R}\!\bra{\phi_{k_{n},-m,\tau}}\hat{L}^{z}\ket{\phi_{k_{n},-m,\tau}}_{R}$.
The same relation holds for the perturbed states including the present $m$-dependent interaction.
This relation implies that reversing the chirality of the crystal flips the sign of the OAM as a function of $k$ with $m \to -m$. This behavior is confirmed numerically using the $\alpha$-inverted Hamiltonian and directly supports the CIOS mechanism discussed here.

Finally, we comment on the connection to CISS phenomena.
Since the interaction $\hat{\mathcal{H}}^{(\mathrm{rot})}_{\mathrm{el\text{-}ph}}$ 
scales as $1/\sqrt{M}$, the effect is stronger in lighter-element systems, 
and the phonon-to-electron OAM transfer increases with temperature. 
Both trends are consistent with experimental observations 
that CISS is active even in light-element systems 
and becomes more pronounced at higher temperatures.
Beyond equilibrium, nonequilibrium chiral-phonon flow may induce 
orbital accumulation, which could be converted into spin polarization 
via spin--orbit coupling at surfaces or interfaces, 
suggesting a microscopic route toward CISS.


In summary, we have developed a symmetry-based microscopic theory of CIOS in helical crystals, in which truly chiral phonons transfer angular momentum selectively to electronic orbital degrees of freedom through a rotational electron--phonon vertex. This framework complements Berry-curvature-based descriptions of the chiral-phonon-induced orbital response~\cite{SatoKato2025,YaoGoMokrousovMurakami2025} by making the local $p$-orbital vertex, nonzero wave vector structure and CAM selection rules explicit. Since CISS is experimentally observed even in 
imperfect helical solids and long chiral molecules where strict 
screw periodicity does not hold, how the present theory survives 
such structural imperfection is left for future work.

\begin{acknowledgments}
We thank Yusuke Kato, Takeo Kato, Takuya Nomoto, Takuya Satoh,
Yoshihiko Togawa, and Hiroshi Yamamoto for fruitful discussions.
T.T. thanks Shiro Komata and Tomokazu Yasuike for inspiring communication.
J.K.\ acknowledges support from JSPS KAKENHI Grant Nos.\ 25H02149,
25K00962, and 23H00091, and from the OML Project grant by the National
Institutes of Natural Sciences (NINS program No.\ OML012301).
This work was also supported by JST ERATO Grant Number JPMJER2503,
Japan.
\end{acknowledgments}


\end{document}